\pdfoutput=1

\documentclass[11pt]{article}

\usepackage[final]{acl}

\usepackage{times}
\usepackage{latexsym}

\usepackage[T1]{fontenc}

\usepackage[utf8]{inputenc}

\usepackage{microtype}

\usepackage{inconsolata}

\usepackage{graphicx}

\usepackage[utf8]{inputenc} 
\usepackage[T1]{fontenc}    
\usepackage{hyperref}       
\usepackage{url}            
\usepackage{booktabs}       
\usepackage{amsfonts}       
\usepackage{nicefrac}       
\usepackage{microtype}      
\usepackage{xcolor}         
\usepackage[colorinlistoftodos,prependcaption,textsize=tiny]{todonotes}
\usepackage{amsthm}
\usepackage{xspace}
\usepackage{wrapfig}
\usepackage{amsmath}
\usepackage{multicol}
\usepackage{multirow}
\usepackage{booktabs}
\usepackage{algorithm}
\usepackage{algorithmic}
\usepackage{stfloats,eucal}
\usepackage{colortbl}
\usepackage{xcolor}
\usepackage{color}
\usepackage{enumitem}
\setlist[itemize]{leftmargin=*}
\definecolor{lm_purple_low}{RGB}{240,240,248}
\usepackage{subcaption}  

\def\ie{{\it i.e.}\xspace}

\def\pP{{\mathcal{P}}}
\def\pX{{\bf {X}}}
\def\pA{{\bf {A}}}
\def\px{{\bf {x}}}
\def\pa{{\bf {a}}}
\def\pC{{\mathcal{C}}}

\newtheorem{definition}{Definition}

\title{From Sentences to Sequences: Rethinking Languages in Biological System}

\author{
 \textbf{Ke Liu \textsuperscript{1}\thanks{Equal contribution.}},
 \textbf{Shuaike Shen \textsuperscript{1}\footnotemark[1]},
 \textbf{Hao Chen \textsuperscript{1}}
\\
 \textsuperscript{1} Zhejiang University
}

\begin{document}
\maketitle
\begin{abstract}

The paradigm of large language models in natural language processing (NLP) has also shown promise in modeling biological languages, including proteins, RNA, and DNA. Both the auto-regressive generation paradigm and evaluation metrics have been transferred from NLP to biological sequence modeling. However, the intrinsic structural correlations in natural and biological languages differ fundamentally. Therefore, we revisit the notion of language in biological systems to better understand how NLP successes can be effectively translated to biological domains. By treating the 3D structure of biomolecules as the semantic content of a sentence and accounting for the strong correlations between residues or bases, we highlight the importance of structural evaluation and demonstrate the applicability of the auto-regressive paradigm in biological language modeling. Code can be found at \href{https://github.com/zjuKeLiu/RiFold}{github.com/zjuKeLiu/RiFold}
\end{abstract}

\section{Introduction}
The paradigm of large language models (LLMs) has demonstrated remarkable success across diverse domains, including natural language processing (NLP)~\cite{gpt2,gpt3,gpt4}, computer vision~\cite{dit}, and biology~\cite{esm2,esm3}. In particular, LLMs have shown strong capabilities in understanding and generating natural language~\cite{gpt2,gpt3,gpt4}. Inspired by their success in NLP, researchers have extended similar generation paradigms and evaluation protocols to biological sequences, such as proteins, RNA, and DNA~\cite{protgpt2,progen2,progen3}. However, the intrinsic structural correlations between natural and biological languages differ fundamentally.

The distinction between biological and natural languages manifests in two primary aspects. \textit{First, contextual dependencies in biological sequences are significantly stronger and more structured than those in natural language}. For instance, base pairing in RNA and hydrogen bonding networks in proteins~\cite{base_pair,rna_pair_example,protein_hydro_bond} give rise to long-range inter-token dependencies that are rare in natural language. \textit{Second, while semantics in NLP are abstract and difficult to quantify, biological sequences have semantics that are physically grounded in their three-dimensional structures, making them directly measurable}. This critical difference implies that evaluation metrics developed for NLP may not be appropriate for biomolecule sequences.

In this work, we take the representative inverse folding problem~\cite{proteinmpnn} as a case study to investigate how structural correlations and evaluation paradigms diverge between biological and natural language modeling. Inverse folding has been widely formulated as a structure-to-sequence translation task, drawing methodological inspiration from neural machine translation~\cite{attention,bench,rdesign}. We adopt this formulation as a foundation to systematically examine these differences and propose biologically appropriate modeling strategies.

From a modeling perspective, we argue that the standard sequential generation paradigm used in NLP is suboptimal for biological sequences. Instead, we demonstrate the effectiveness of stochastic-order generation. Unlike natural language, where adjacent tokens tend to be semantically related, biomolecule sequences often contain long-range dependencies due to physical interactions such as base pairing and hydrogen bonds. In particular, distant tokens in the 1D sequence may be spatially proximal in the 3D structure, where positional proximity generally aligns with semantic dependency. This is in sharp contrast to the natural sentence. Moreover, while replacing a word with its synonym typically preserves semantics in NLP, substituting even a single residue in a protein or base in RNA can lead to complete structural collapse. We find that stochastic-order decoding better captures such complex dependencies and preserves structural fidelity.

For evaluation, we advocate for structure-based metrics over sequence-based ones. Traditional NLP metrics, such as BLEU~\cite{bleu} and ROUGE~\cite{rouge}, measure the similarity between predicted sentences and ground truth under the assumption that semantic meaning is encoded in the token sequence. In biological systems, however, even minor sequence perturbations can result in radically different 3D conformations, and the semantics are closely tied to their 3D structures.
Thus, structure recovery metrics, such as TM-score, RMSD, and energy, are more appropriate to access semantic fidelity in biomolecule inverse folding. 
In addition, the semantic similarity, which is hard to measure in NLP, is physically the 3D structure of biomolecules. Therefore, by treating the 3D structure as the semantic representation of a sequence, we enable a more meaningful evaluation of semantic similarity in biological sequences.

To the best of our knowledge, this is the first work to analyze the differences between natural and biological language. The main contributions of our work can be summarized as follows:

\begin{figure*}[tb!]
    \vspace{-1em}
    \centering
    \includegraphics[width=0.9\linewidth]{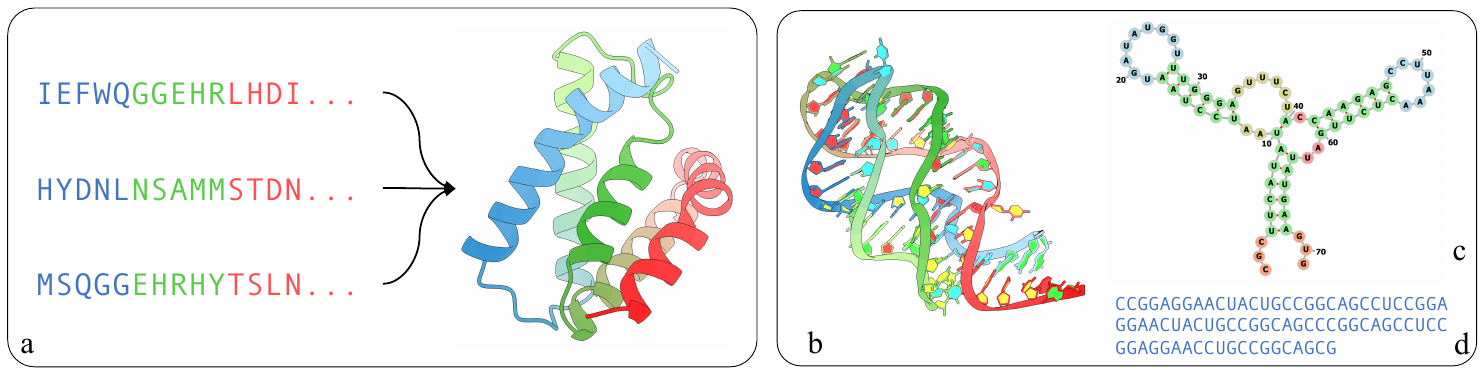}
    \caption{Structures of protein and RNA. (a) protein structure and sequence. One protein structure corresponds to multiple sequences. (b) RNA tertiary structure. Base pairs exist in RNA, which is different from protein. (c) RNA secondary structure. (d) RNA primary structure, \ie, RNA sequence. }
    \label{fig:structure}
\end{figure*}

\begin{itemize}

\item We provide an in-depth analysis of the difference between the biological and natural language. We demonstrate that the stochastic-order generation paradigm works better than sequential-order generation for biomolecule sequences on the inverse folding task.

\item We propose a more comprehensive evaluation pipeline for biomolecule inverse folding problem, which can better evaluate the high-level semantic meaning of biological language.

\item We explore the gap between structure and sequence recovery. Empirical results demonstrate that these recoveries are related but not consistent, indicating the token-level recovery does not align with the high semantic level similarity, which is different from natural language.
\end{itemize}

\section{Related Works}

\subsection{Sequential-order and Stochastic-order Generation}
Sequential autoregressive models find wide application in tasks such as image generation \cite{image-gpt,dit} and natural language processing \citep{gpt,gpt2,gpt3,gpt4}. In sequential generation, outputs are produced strictly left-to-right. In contrast, stochastic-order generation allows emitting tokens at arbitrary positions without fixed ordering constraints.

\subsection{biomolecule Inverse Folding}
The task of biomolecule inverse folding is to translate the given structure into corresponding sequence. Specifically, predicting amino acid sequence for the given protein structure \cite{proteinmpnn,pifold,lmdesign} or generating the sequence of ribonucleic acids corresponding to a specified RNA tertiary structure \cite{rdesign}, adhering to the principle of base pairing \cite{base_pair}. 
Due to the long-range inter-token dependencies mentioned previously \cite{hydrogen_bond,base_pair}, Sequential-order generation methods typically focus only on local contextual information surrounding the currently generated token, which will lead to the failure of recovering the dependencies.

\subsection{Evaluation Metrics for Sequence Data}
When fine-tuning large language models, the most common metrics to evaluate the similarity between predicted sequences and ground truth are BLEU score \cite{bleu} and ROUGE score \cite{rouge}. Specifically, BLEU score measures the precision of n-grams between the machine-translated output and human reference translations, and ROUGE is a set of metrics primarily used for evaluating automatic summarization. For machine translation tasks, researchers usually utilize multi-character level metrics to evaluate context consistency \cite{bleu,rouge,meteor,ter,chrF} or embeddings from pre-trained Language Models \cite{bertscore,comet} to evaluate semantic similarity. However, the biomolecule inverse folding task, which can also be treated as a translation task from 3D structure to 1D sequence, only takes sequence recovery into consideration. This does not take into account the relationships between tokens and the semantic consistency.

\begin{figure*}[ht!]
    \vspace{-2.1em}
    \centering
    \includegraphics[width=0.9\linewidth]{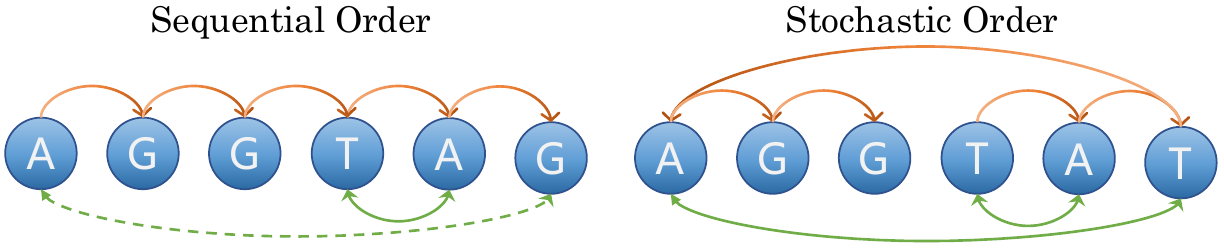}
    \caption{In sequential-order generation, tokens are generated from left to right and tokens are allowed to be decoded in any order without any constraints in stochastic order. \textcolor{orange}{Orange arrows} denote the generating order and \textcolor{green}{green arrows} indicate the interactions such as base pairing in RNA.
    In this work, the generation order is determined based on the confidence of each time step.}
    \label{fig:decoding_order}
    \vspace{-1em}
\end{figure*}

\section{Preliminaries and Background}
\subsection{Sequential-order and Stochastic-order Generation}
\paragraph{Sequential Generation}
As Fig.~\ref{fig:decoding_order} shows, output tokens are strictly produced left-to-right, with each timestep constrained to generate only the current position's token:
\begin{equation}
    \label{pre:sequential}
    P(y_t | y_{<t}, x), \quad t \in \{1,2,...,n\},
\end{equation}
where $x$ is the input sequence, $y = (y_1, y_2, ..., y_n)$ is the output sequence and $y_{<t} = (y_1, ..., y_{t-1})$ represents the tokens generated before the current position. The generation order is fixed to $ 1 \rightarrow 2 \rightarrow \cdots \rightarrow n$.

\vspace{-0.5em}
\paragraph{Stochastic Generation}
In contrast to sequential approaches, stochastic-order generation permits token production at any valid position:
\begin{equation}
    \label{pre:stochastic}
    P(y_{p_t} | y_{S_{<t}}, x), \quad p_t \in \mathcal{P}\setminus S_{<t},
\end{equation}
where $\mathcal{P} = \{1,2,...,n\}$ is the set of all position indices, $S = (p_1, p_2, ..., p_n)$ is the sequence of generated positions (permutation) and $S_{<t} = \{p_1, ..., p_{t-1}\}$ represents the position generated in the previous $ t-1 $ step. Each time $p_t$ is selected from the remaining positions $\mathcal{P}\setminus S_{<t}$ to generate.

\vspace{-0.5em}
\subsection{biomolecule Inverse Folding Problem}
The biomolecule inverse folding problem is treated as a structure-sequence translation problem \cite{proteinmpnn}. 
A biomolecule $\pP = \{\pA, \pX\}$ consists of its sequence $\pA = [ \pa_1, \pa_2, \ldots , \pa_n ]$ and backbone structure $\pX = [ \px_1,\px_2, \ldots , \px_n ]^T \in \mathbb{R}^{n \times 3}$, where $n$ denotes the number of bases in a biomolecule. In this work, we focus on proteins and RNA. For proteins, the sequence refers to the amino acid sequence, and $\pa_i \in \pC^{20}$ denotes the type of $i$-th amino acids, where $\pC$ is a set of 20 genetically-encoded amino acids. For RNA, the sequence is ribonucleic acids sequence. $\pa_i \in \pC^{4}$ and $\px_i \in \mathbb{R}^3 $ denote types and positions of $i$-th ribonucleic acids. The inverse folding problem aims to design sequences based on specified tertiary structures, which can be defined as:

\vspace{-0.5em}
\begin{definition}[Biomolecule inverse folding]
    Given the structure $\pX$ of biomolecules, the biomolecule inverse folding seeks to translate the structure to corresponding sequence, \ie, $\hat{\pA} = f (\pX)$.
\end{definition}
\vspace{-0.5em}

\subsection{Protein and RNA Structures}
One biomolecule tertiary structure corresponds to multiple sequences as shown in Fig.~\ref{fig:structure}(a) \citep{str_multi_seq}. A slight difference in the critical position of a sequence may result in a totally different tertiary structure. Compared to proteins, RNA sequences exhibit stronger internal dependencies due to well-defined base pairing rules, as illustrated in Fig.~\ref{fig:structure}(b)(c). Specifically, guanine (G) typically pairs with cytosine (C), and adenine (A) pairs with uracil (U), the RNA counterpart of thymine~\citep{rna_pair} as shown in Fig.~\ref{fig:structure}(d), where indicates tokens distant in the 1D sequence may be spatially proximal in the 3D structure for biomolecules.

\subsection{Evaluation Metrics}
Native sequence recovery (NSR) is a commonly used metric in the inverse folding problem:
\begin{equation}
    \label{eq:nsr}
    {\rm NSR} = \frac{1}{|\pA|} \sum_{i=1}^{|\pA|} \delta(\pa_i, \hat{\pa}_i),
\end{equation}
where $\hat{\pa}_i$ is the $i$-th prediction of the model. $\delta$ indicates the Kronecker delta function, which takes the value of $1$ when its arguments are the same and $0$ when they are different. 
However, the aim of inverse folding is to design a sequence that can be folded into the desired tertiary structure, which is similar to generating sentences with specific semantics in NLP.  Although a minor synonym substitution has little effect on semantics in NLP, a slight difference in sequences at critical positions may result in a totally different structure. Therefore, NSR from NLP is not an appropriate evaluation metric for the inverse folding problem.

\begin{figure}[t!]
    \centering
    \includegraphics[width=0.9\linewidth]{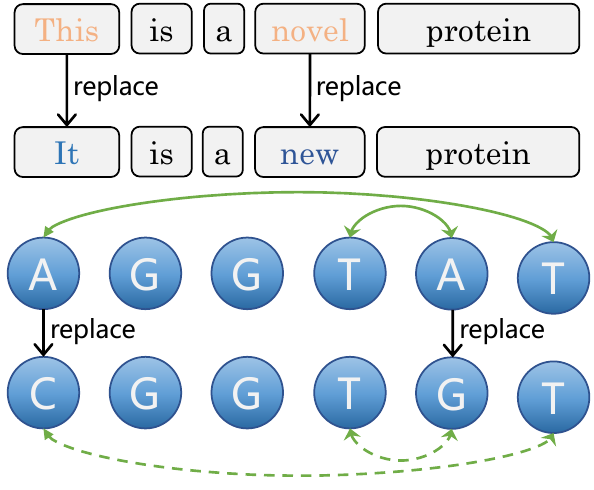}
    \vspace{-1em}
    \caption{Replacing words with synonyms will not lead to a significant change in the meaning of the sentence. In contrast, in RNA sequences, substitution of any single nucleotide may disrupt critical base pairing interactions. \textcolor{green}{Green arrows} denote base pairing in RNA structures.}
    \label{fig:semantic}
\end{figure}

\section{Differences Between Natural and Biomolecular Languages}

\subsection{Long-Range Inter-Token Dependencies}
Natural languages generally follow a proximity principle in syntax, wherein the likelihood of a strong dependency relation typically decreases as the distance between words increases. This observation helps explain the prevalence of sequential autoregressive paradigms in popular language models such as the GPT series~\cite{gpt,gpt2,gpt3,gpt4}; by maintaining local coherence within a generation window, these models effectively achieve global fluency in generated text.

In contrast, biomolecular sequences often exhibit long-range inter-token dependencies—for example, RNA base pairing or protein residue interactions (Fig.\ref{fig:structure}). In such cases, enforcing only local coherence is insufficient. As shown in Fig.\ref{fig:decoding_order}, strict sequential-order generation fails to capture these long-range dependencies, leading to incorrect base-pairing predictions. A stochastic generation order can better preserve such dependencies, as tokens with strong interrelationships are more likely to be generated in adjacent timesteps. To test this hypothesis, we evaluate both generation paradigms on a biomolecular inverse folding task. We find that our RNA inverse folding model outperforms existing baselines, as detailed in Sec.~\ref{sec:rna_if}.

\subsection{Semantic Representation}
In natural language, semantics refers to the meaning conveyed by linguistic expressions and emerges from interactions among words. Critically, these semantics are abstract and lack any physical form. In contrast, the semantics of a biomolecular sequence are concrete, directly grounded in its three-dimensional structure and energetic properties. Each token in a biomolecule contributes directly to the global semantic state (i.e., the folded structure and stability), making such sequences highly sensitive to single-token perturbations. As illustrated in Fig.~\ref{fig:semantic}, whereas a synonym substitution in a sentence typically preserves its meaning, replacing a single RNA nucleotide can disrupt base pairing and cause the structure to collapse.

\subsection{Evaluation Pipeline}
Given these semantic differences, more comprehensive metrics are required to evaluate biomolecular inverse folding. Prior work often considers only native sequence recovery, which by itself fails to adequately assess the preservation of structural semantics. To address this limitation, we propose a structure-aware evaluation pipeline that incorporates \textbf{structure recovery} (TM-score and RMSD), \textbf{energy}, and \textbf{sequence recovery}.

\begin{figure}[t!]
    \centering
    \includegraphics[width=0.8\linewidth]{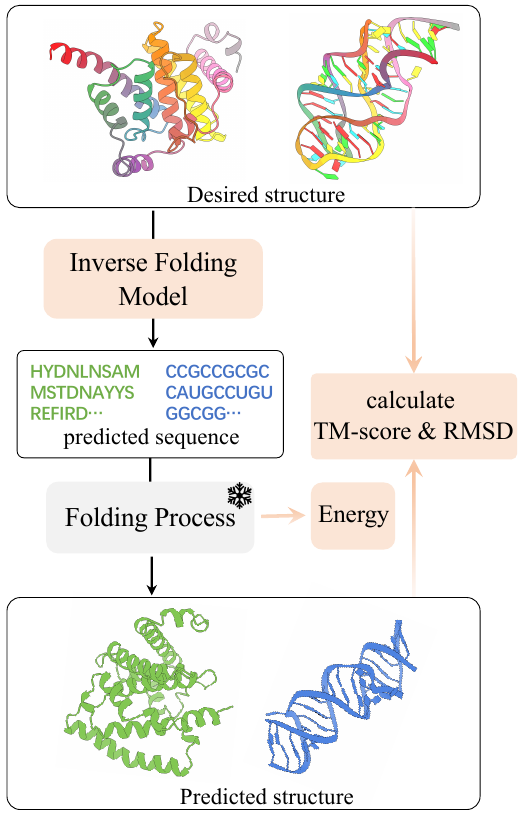}
    \vspace{-0.8em}
    \caption{Evaluation workflow for biomolecule inverse folding.}
    \label{fig:evaluation}
\end{figure}

Fig.\ref{fig:evaluation} illustrates this process. Given a target tertiary structure $\pX$, our model generates a candidate sequence $\hat{\pA} = f(\pX)$, which is then folded into a predicted structure $\hat{\pX}$. We then compare $\hat{\pX}$ against $\pX$ using TM-score and RMSD\cite{se3_diffusion}. According to standard criteria, we consider the structure successfully recovered if $\text{TM-score} > 0.5$ and $\text{RMSD} < 2$~\AA.

We follow domain-specific conventions when computing these metrics: for proteins, structural similarity is assessed using C$_\alpha$ atoms~\cite{rfdiffusion,se3_diffusion}, whereas for RNA we use the C3$'$ and C4$'$ atoms of the sugar backbone. Additionally, to quantify RNA stability, we introduce an \textbf{energy} metric computed using E2Efold~\cite{RNAFold2}; lower energy indicates a more stable (and thus more plausible) structure.

\section{Experiments \& Discussion}

\subsection{RNA Inverse Folding}
\label{sec:rna_if}
\subsubsection{Setup}
\paragraph{Dataset.} This study leverages two datasets for RNA inverse folding, namely, \textbf{RNAsolo} \citep{rnasolo} and \textbf{RNA-Puzzles} \citep{puzzle}, following the precedent set by previous research \citep{rdesign}. These datasets encompass the RNA tertiary structure and sequence. The data splitting pursued parallels the methodology adopted by Tan et al. \citep{rdesign}.

\paragraph{Evaluation metrics.} Our principal assessment tools for RNA inverse folding include structure recovery metrics such as \textbf{TM-score} and \textbf{RMSD}, and sequence recovery metrics comprising native structure recovery (\textbf{NSR}) and Macro-F1 (\textbf{M-F1}) score, following the approach utilized in \citep{rdesign}. The NSR is described as Eq.~\ref{eq:nsr} and Macro-F1 score is computed as described in \citep{rdesign}:
\begin{equation*}
    \text{M-F1} = \frac{1}{|T|} \sum_{t\in\{{\rm A, U, C, G}\}}^{T} 2\times \frac{P_t \times R_t}{P_t + R_t},
\end{equation*}
where $P_t$ and $R_t$ are the precision and recall of token $c$ respectively.
\paragraph{Compared approaches.} Following Tan et al. \citep{rdesign}, we compare our RiFold with sequence-based models (\textit{SeqRNN} and \textit{SeqLSTM}), tertiary structure-based models (\textit{RDesign} \citep{rdesign}, \textit{StructMLP}, \textit{StructGNN}, \textit{GraphTrans} \citep{StructGNN}, and \textit{PiFold} \citep{pifold}), and secondary structure-based models (\textit{MCTS-RNA} \citep{MCTS}, \textit{LEARNA} \citep{LEARNA}, \textit{eM2dRNAs} \citep{eM2dRNAs}, and \textit{aRNAque} \citep{aRNAque}).

\begin{table*}[ht!]
\small
\vspace{-0.5em}
\centering
\setlength{\tabcolsep}{4.5mm}{
\begin{tabular}{ccccccccccc}
\toprule
\multirow{2}{*}{Method} & \multicolumn{4}{c}{Native Sequence Recovery (\%) $\uparrow$}\\
& Short   & Medium   & Long  & All \\
\midrule
SeqRNN (h=128)          & 26.52$\pm$1.07 & 24.86$\pm$0.82 & 27.31$\pm$0.41 & 26.23$\pm$0.87 \\
SeqRNN (h=256)          & 27.61$\pm$1.85 & 27.16$\pm$0.63 & 28.71$\pm$0.14 & 28.24$\pm$0.46 \\
SeqLSTM (h=128)         & 23.48$\pm$1.07 & 26.32$\pm$0.05 & 26.78$\pm$1.12 & 24.70$\pm$0.64 \\
SeqLSTM (h=256)         & 25.00$\pm$0.00 & 26.89$\pm$0.35 & 28.55$\pm$0.13 & 26.93$\pm$0.93 \\
StructMLP               & 25.72$\pm$0.51 & 25.03$\pm$1.39 & 25.38$\pm$1.89 & 25.35$\pm$0.25 \\
StructGNN               & 27.55$\pm$0.94 & 28.78$\pm$0.87 & 28.23$\pm$1.95 & 28.23$\pm$0.71 \\
GraphTrans              & 26.15$\pm$0.93 & 23.78$\pm$1.11 & 23.80$\pm$1.69 & 24.73$\pm$0.93 \\
PiFold                  & 24.81$\pm$2.01 & 25.90$\pm$1.56 & 23.55$\pm$4.13 & 24.48$\pm$1.13 \\
RDesign                 & 37.22$\pm$1.14 & 44.89$\pm$1.67 & 43.06$\pm$0.08 & 41.53$\pm$0.38 \\
\cellcolor{lm_purple_low}RiFold                  & \cellcolor{lm_purple_low}\textbf{41.23}$\pm$2.10 & \cellcolor{lm_purple_low}\textbf{45.23}$\pm$1.43 & \cellcolor{lm_purple_low}\textbf{43.88}$\pm$0.53 & \cellcolor{lm_purple_low}\textbf{43.04}$\pm$1.02 \\
\bottomrule
\end{tabular}}
\caption{The sequence recovery on RNAsolo dataset. The best results are highlighted in bold.}
\label{tab:sequence_recovery}
\end{table*}

\begin{table*}[t!]
\small
\centering
\vspace{-0.5em}
\setlength{\tabcolsep}{4.5mm}{
\begin{tabular}{ccccccccccc}
\toprule
\multirow{2}{*}{Method} & \multicolumn{4}{c}{Macro F1 ($\times$100) $\uparrow$} \\
& Short   & Medium   & Long  & All \\
\midrule
SeqRNN (h=128)          & 17.22$\pm$1.69 & 17.20$\pm$1.91 & 8.44$\pm$2.70  & 17.74$\pm$1.59 \\
SeqRNN (h=256)          & 12.54$\pm$2.94 & 13.64$\pm$5.24 & 8.85$\pm$2.41  & 13.64$\pm$2.69 \\
SeqLSTM (h=128)         & 9.89$\pm$0.57  & 10.44$\pm$1.42 & 10.71$\pm$2.53 & 10.28$\pm$0.61 \\
SeqLSTM (h=256)         & 9.26$\pm$1.16  & 9.48$\pm$0.74  & 7.14$\pm$0.00  & 10.93$\pm$0.15 \\
StructMLP               & 17.46$\pm$2.39 & 18.57$\pm$3.45 & 17.53$\pm$8.43 & 18.88$\pm$2.50 \\
StructGNN               & 24.01$\pm$3.62 & 22.15$\pm$4.67 & 26.05$\pm$6.43 & 24.87$\pm$1.65 \\
GraphTrans              & 16.34$\pm$2.67 & 16.39$\pm$4.74 & 18.67$\pm$7.16 & 17.18$\pm$3.81 \\
PiFold                  & 17.48$\pm$2.24 & 18.10$\pm$6.76 & 14.06$\pm$3.53 & 17.45$\pm$1.33 \\
RDesign                 & 38.25$\pm$3.06 & 40.41$\pm$1.27 & 41.48$\pm$0.91 & 40.89$\pm$0.49 \\
\cellcolor{lm_purple_low}RiFold                  & \cellcolor{lm_purple_low}\textbf{39.87}$\pm$1.41 & \cellcolor{lm_purple_low}\textbf{45.13}$\pm$1.55 & \cellcolor{lm_purple_low}\textbf{42.82}$\pm$0.37 & \cellcolor{lm_purple_low}\textbf{43.17}$\pm$0.75 \\
\bottomrule 
\end{tabular}}
\caption{The Macro-F1 on the RNAsolo dataset. The best results are highlighted in bold. }
\label{tab:f1}
\end{table*}

\subsubsection{Experimental results}

\paragraph{Sequence recovery.} We first compare our stochastic-order based RiFold with other models based on previous sequence recovery metrics. The results on sequence recovery are shown in Table~\ref{tab:sequence_recovery} and the results on Macro-F1 are shown in Table~\ref{tab:f1}. The short, medium, and long indicate the RNA with lengths of 0 to 50, 50 to 100, and more than 100 acids. Empirical results demonstrate that our RiFold with stochastic-order generation outperforms previous works. RiFold achieves a 3.64\% improvement in sequence recovery and a 5.57\% improvement in Macro-F1 over the previous State-Of-The-Art (SoTA) model, RDesign. Our RiFold outperforms RDesign for two reasons: (1) The strong context correlation is better maintained by our RiFold. On the samples with base pairing, RiFold is much better than RDesign, as shown in Table~\ref{tab:pair_result}. (2) High confidence for each predicted token. The average confidence of our RiFold achieves 0.9215, while the average confidence for RDesign is only 0.4356, which means that RiFold is not certain about its prediction. This is caused by the problem that one structure corresponds to multiple sequences, and our stochastic-order model is able to maintain the structure consistency.

\begin{table}[ht]
\small
\centering
\begin{tabular}{c cc cc}
\toprule
 Metric & Method & Short & Medium & Long \\
 \midrule
 \multicolumn{2}{c}{\# Samples} & 52 & 58 & 26 \\
 \multirow{2}{*}{M-F1 $\uparrow$} & RDesign & 38.75 & 44.19 & 41.23 \\
 & \cellcolor{lm_purple_low}RiFold & \cellcolor{lm_purple_low}\textbf{44.22} & \cellcolor{lm_purple_low}\textbf{46.67} & \cellcolor{lm_purple_low}\textbf{42.95} \\
 \multirow{2}{*}{NSR $\uparrow$} & RDesign & 39.84 & 45.42 & 43.16 \\
 & \cellcolor{lm_purple_low}RiFold & \cellcolor{lm_purple_low}\textbf{46.67} & \cellcolor{lm_purple_low}\textbf{46.44} & \cellcolor{lm_purple_low}\textbf{44.07} \\
\bottomrule
\end{tabular}%
\caption{Experimental results on the samples with base pairs in solo RNA dataset.}\label{tab:pair_result}
\end{table}

\begin{figure*}[t!]
    \centering
    \includegraphics[width=1\linewidth]{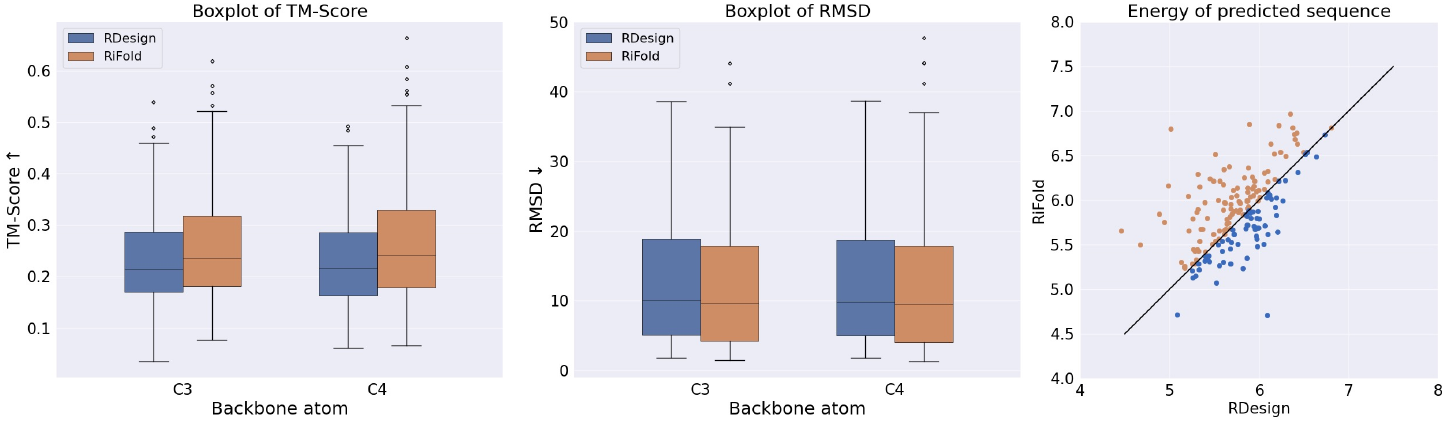}
    \vspace{-2em}
    \caption{Structure recovery and energy comparison between RiFold and RDesign. C3 and C4 indicate the results are calculated with carbon 3 and 4 as the backbone. Left: Boxplot of TM-score. Middle: Boxplot of RMSD. Right: Scatterplot of energy. Horizontal and vertical coordinates are energies (log-) of sequences predicted by RDesign and RiFold respectively. Larger is better. The points over the black line indicate that RiFold outperforms RDesign.}
    \label{fig:energy}
\end{figure*}

\paragraph{Structure recovery.} We conducted an evaluation of RiFold alongside the prior SOTA method, RDesign, employing more appropriate metrics; energy and structure recovery, which encompasses RMSD and TM-score as shown in Table~\ref{tab:structure_recovery}. In particular, RiFold outperforms RDesign in all three metrics, as depicted in Fig.~\ref{fig:energy}. RiFold achieves improvements of 13.88\% and 5.86\% in the average TM-score and RMSD, respectively, compared to RDesign. Indeed, 60.22\% of the sequences predicted by RiFold achieved a lower energy than RDesign. Moreover, computing structure recovery based on either Carbon-3 or Carbon-4 atoms results in only minor differences.

\begin{table}[t!]
\small
    \centering
    \setlength{\tabcolsep}{4pt} 
\renewcommand{\arraystretch}{1.1} 
\begin{tabular}{
    >{\centering\arraybackslash}p{1.2cm}  
    >{\centering\arraybackslash}p{0.5cm}  
    >{\centering\arraybackslash}p{1.1cm}  
    >{\centering\arraybackslash}p{1.1cm}  
    >{\centering\arraybackslash}p{1.1cm}  
    >{\centering\arraybackslash}p{1.1cm}  
}
\toprule
    \multirow{2}{*}{Method}& & \multicolumn{2}{c}{Mean} & \multicolumn{2}{c}{Median} \\
     \cmidrule{3-4}  \cmidrule{5-6} 
    & & RDesign & \cellcolor{lm_purple_low} RiFold & Rdesign & \cellcolor{lm_purple_low} RiFold \\
     \hline
    \multirow{2}{*}{TM-Score $\uparrow$} & C3 &0.2315 & \cellcolor{lm_purple_low}\textbf{0.2580}& 0.2148 & \cellcolor{lm_purple_low}\textbf{0.2365}\\
     & C4 &  0.2317 & \cellcolor{lm_purple_low}\textbf{0.2695} &0.2165  & \cellcolor{lm_purple_low}\textbf{0.2407} \\
     \hline
    \multirow{2}{*}{RMSD $\downarrow$}  & C3 &12.8416 & \cellcolor{lm_purple_low}\textbf{12.0581} & 9.9956 & \cellcolor{lm_purple_low}\textbf{9.5949}\\
     & C4 &  12.7414 &\cellcolor{lm_purple_low}\textbf{12.0386} & 9.8600 & \cellcolor{lm_purple_low}\textbf{9.5318}\\
     \hline
     Energy $\uparrow$ & & 5.7646 &\cellcolor{lm_purple_low}\textbf{5.8757}  & 5.7762 & \cellcolor{lm_purple_low}\textbf{5.8686} \\
    \bottomrule
    \end{tabular}
    \caption{The structure recovery on the RNAsolo dataset. Energy(log-), RMSD(\AA).}
    \label{tab:structure_recovery}
    \vspace{-0.5em}
\end{table}

\paragraph{Ablation study.} We conduct ablation studies to verify the effectiveness of RiFold. The beam search and stochastic order decoding do work in RiFold, as shown in Table~\ref{tab:ablation}. With beam search, RiFold achieves improvements of 1.44\% and 0.42\% on macro-F1 and sequence recovery. With stochastic order decoding, RiFold achieves improvements of 3.67\% and 3.28\% on macro-F1 and sequence recovery. Besides, stochastic order decoding improves the performance of RiFold especially on short RNA sequences. The average improvements of macro-F1 and sequence recovery on short RNA sequences are 6.94\% and 6.76\% respectively. Order decoding takes the decoding order, \ie, the position of tokens in the sequence, as important information, which should not be considered in the biomolecule inverse folding problem. Stochastic order decoding removes the dependency on decoding order by decoding the tokens in a random order. More results for beam search can be found in Appendix.~\ref{app:beamsearch}.

\begin{table}[t!]
\small 
    \centering
    \begin{tabular}{lcccc}
        \toprule
        \textbf{Method} & \textbf{Short} & \textbf{Medium} & \textbf{Long} & \textbf{All} \\
        \midrule
        & \multicolumn{4}{c}{\textit{Macro-F1 ($\times$ 100) $\uparrow$}} \\
        \cmidrule{2-5}
        w/o SO & 38.39 & 44.87 & 40.04 & 41.64 \\
        w/o BS & 39.78 & 43.96 & 42.35 & 42.56 \\
        \rowcolor{lm_purple_low}
        RiFold & \textbf{39.87} & \textbf{45.13} & \textbf{42.82} & \textbf{43.17} \\
        \midrule
        & \multicolumn{4}{c}{\textit{Sequence Recovery (\%) $\uparrow$}} \\
        \cmidrule{2-5}
        w/o SO & 41.17 & 44.96 & 41.10 & 41.67 \\
        w/o BS & 41.18 & 44.36 & 43.05 & 42.86 \\
        \rowcolor{lm_purple_low}
        RiFold & \textbf{41.23} & \textbf{45.23} & \textbf{43.88} & \textbf{43.04} \\
        \bottomrule
    \end{tabular}
    \caption{Ablation study of \textsc{RiFold}. SO and BS indicate stochastic order decoding and beam search, respectively.}
    \label{tab:ablation}
    \vspace{-0.5em}
\end{table}

\begin{table}[ht!]
    \small
    \centering
    \begin{tabular}{p{1.6cm}<{\centering}p{1.4cm}<{\centering}p{1.4cm}<{\centering}p{1.4cm}<{\centering}}
        \toprule
        & \multicolumn{3}{c}{Structure Recovery} \\
        \cmidrule{2-4}
        \textbf{Method} & TM-score > 0.5 (\%) $\uparrow$ & RMSD < 2 (\%) $\uparrow$ & \cellcolor{lm_purple_low} TM > 0.5 \& RMSD < 2 (\%) $\uparrow$ \\
        \midrule
        KWDesign       & 89.10 & 60.59 & \cellcolor{lm_purple_low}60.59 \\
        PiFold         & 90.93 & 59.44 & \cellcolor{lm_purple_low}59.44 \\
        \cellcolor{lm_purple_low}PiFold-AR     & \cellcolor{lm_purple_low}89.29 & \cellcolor{lm_purple_low}60.98 & \cellcolor{lm_purple_low}\underline{60.98} \\
        \cellcolor{lm_purple_low}ProteinMPNN   & \cellcolor{lm_purple_low}90.88 & \cellcolor{lm_purple_low}61.04 & \cellcolor{lm_purple_low}\textbf{61.04} \\
        \bottomrule
    \end{tabular}
    
    \vspace{0.8em}

    \begin{tabular}{p{1.6cm}<{\centering}p{1.8cm}<{\centering}p{1.8cm}<{\centering}}
        \toprule
        & \multicolumn{2}{c}{Sequence Recovery} \\
        \cmidrule{2-3}
        \textbf{Method} & Perplexity $\downarrow$ & NSR (\%) $\uparrow$ \\
        \midrule
        KWDesign       & \textbf{4.42} & \textbf{60.13} \\
        PiFold         & \underline{4.58} & \underline{52.17} \\
        \cellcolor{lm_purple_low}PiFold-AR     & \cellcolor{lm_purple_low}4.90 & \cellcolor{lm_purple_low}51.41 \\
        \cellcolor{lm_purple_low}ProteinMPNN   & \cellcolor{lm_purple_low}4.61 & \cellcolor{lm_purple_low}45.96 \\
        \bottomrule
    \end{tabular}

    \caption{Structure and sequence recovery results on the CATH 4.2 protein benchmark. The best and second-best results are marked in \textbf{bold} and \underline{underline}, respectively. Shadowed rows indicate autoregressive methods. The key metric for structural quality is TM-score $>$ 0.5 \& RMSD $<$ 2\AA.}
    \label{tab:protein}
\end{table}

\paragraph{Generalization of RiFold.} To demonstrate the generalization of RiFold, we conducted additional evaluations of our RiFold on the RNA-Puzzles dataset \citep{puzzle}, in accordance with Tan et al. \citep{rdesign}. All models are trained using the RNAsolo dataset and subsequently evaluated on the RNA-Puzzles dataset. RiFold surpasses the performance of all previous models, illustrating a strong capacity for generalization, as shown in Table~\ref{tab:generalization}.

\begin{table}[h!]
\centering
\small
\setlength{\tabcolsep}{1mm}{
\begin{tabular}{@{}lcc@{}}
\toprule
Method & Sequence Recovery (\%) $\uparrow$ & Macro F1 ($\times$100) $\uparrow$ \\
\midrule
SeqRNN & 31.25$\pm$0.72 & 13.24$\pm$1.25 \\
SeqLSTM & 31.62$\pm$0.20 & 12.22$\pm$0.21 \\
StructMLP & 24.22$\pm$1.28 & 16.40$\pm$3.28 \\
StructGNN  & 27.96$\pm$3.08 & 22.76$\pm$3.19 \\
GraphTrans & 22.21$\pm$2.98 & 17.04$\pm$5.36 \\
PiFold & 23.78$\pm$6.52 & 16.20$\pm$3.49 \\
MCTS-RNA  & 32.06$\pm$1.87 & 24.12$\pm$3.47 \\
LEARNA & 30.94$\pm$4.15 & 22.75$\pm$ 1.17 \\
aRNAque & 31.07$\pm$2.32 & 23.30$\pm$1.65 \\
eM2dRNAs & 37.10$\pm$3.24 & 26.91$\pm$2.32 \\
RDesign  & 50.12$\pm$1.07 & 49.24$\pm$1.07 \\
\cellcolor{lm_purple_low}RiFold  & \cellcolor{lm_purple_low}\textbf{56.51}$\pm$0.60 & \cellcolor{lm_purple_low}\textbf{59.32}$\pm$0.22 \\
\bottomrule
\end{tabular}}
\caption{The overall sequence recovery and Macro-F1 scores on the RNA-Puzzles dataset.}
\label{tab:external_result_ori}
\label{tab:generalization}
\end{table}

\subsection{Protein Inverse Folding}
\subsubsection{Setup}
\paragraph{Dataset.} In this work, the \textbf{CATH} dataset \citep{cath}, widely adopted in protein inverse folding, is employed. We follow the data splitting of preceding works \citep{StructGNN, pifold}, in which proteins are divided according to the CATH topology principles, giving rise to a structure of 18024 proteins for training, 608 for validation, and 1120 for testing. 

\paragraph{Evaluation metrics.} We mainly employ structure recovery metrics such as \textbf{TM-score} and \textbf{RMSD}, and sequence recovery metrics comprising perplexity and native structure recovery (\textbf{NSR}) for protein inverse folding evaluation in this work. The structure recovery is the same as the metrics for RNA and the sequence recovery follows previous works \citep{pifold,bench}.

\paragraph{Compared approaches.} We mainly compare autoregressive methods (PiFold-AR and ProteinMPNN \citep{proteinmpnn}) with non-autoregressive methods (StructGNN \citep{StructGNN}, GraphTrans \citep{StructGNN}, GCA \citep{GCA}, GVP \citep{GVP}, AlphaDesign \citep{alphadesign}, PiFold \citep{pifold}, KWDesign \cite{kwdesign}). PiFold-AR is implemented with the encoder of PiFold and a stochastic order decoding autoregressive decoder.

\begin{figure*}[ht!]
    \centering
    \begin{minipage}[b]{0.48\textwidth}
    \centering
        \includegraphics[width=0.95\textwidth]{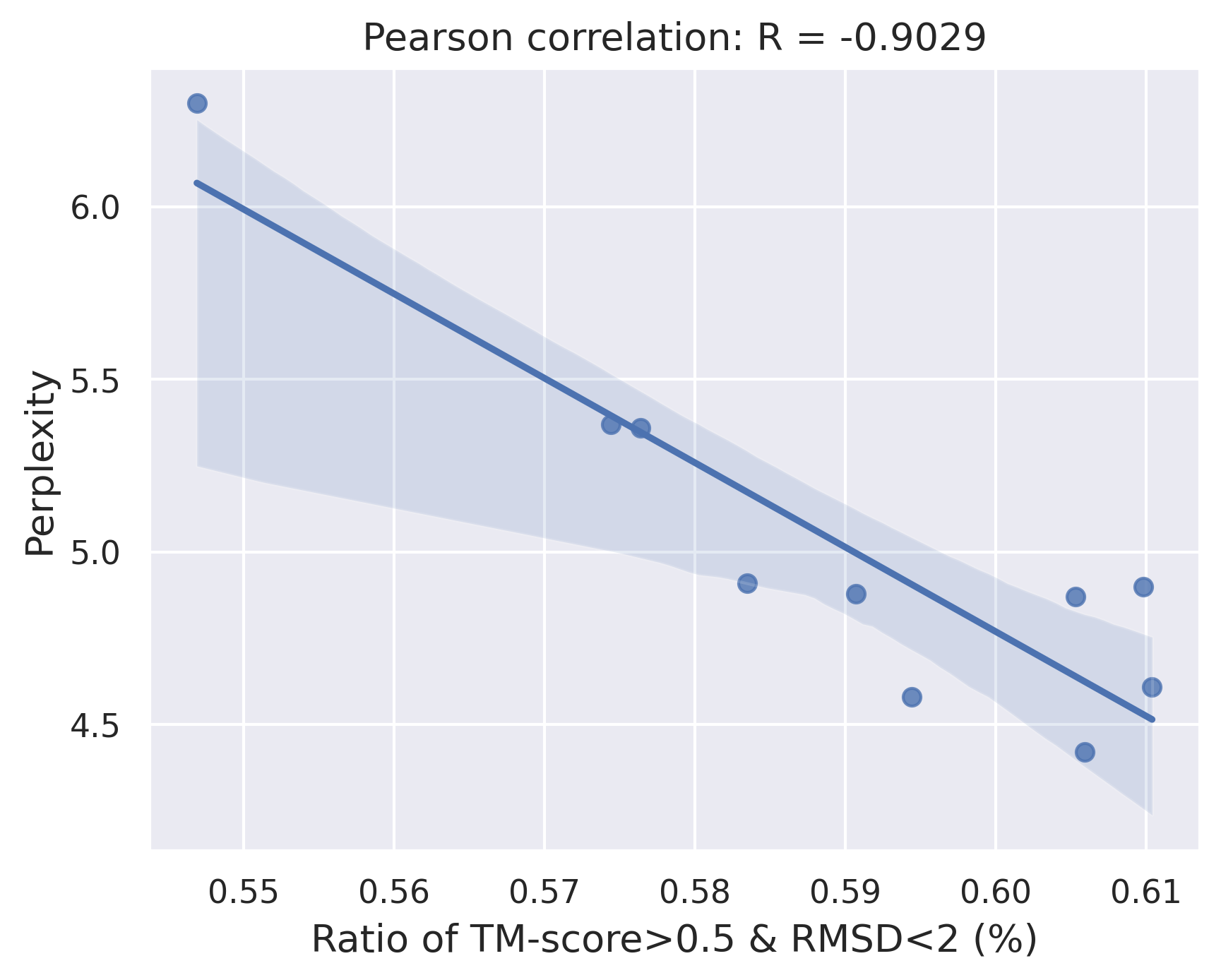}
    \end{minipage}
    \hfill
    \begin{minipage}[b]{0.48\textwidth}
    \centering
        \includegraphics[width=0.95\textwidth]{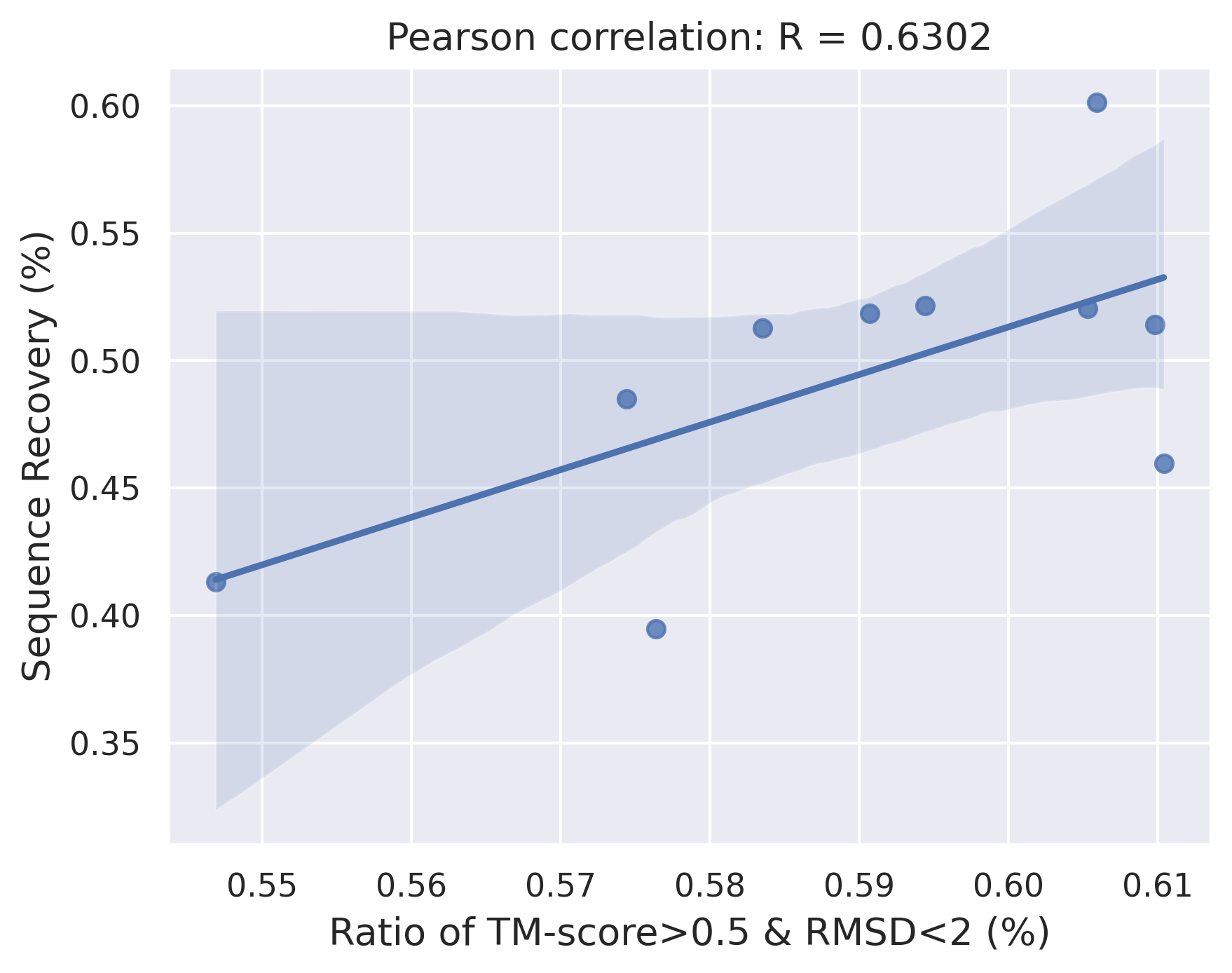}
    \end{minipage}
    \vspace{-1em}
    \caption{The correlation between structure and sequence recovery.}
    \label{fig:corr}
     \vspace{-1em}
\end{figure*}

\subsubsection{Sequence and Structure Recovery Gap}
We evaluate autoregressive and non-autoregressive methods for protein inverse folding on the metrics of structure recovery and sequence recovery. 
Autoregressive methods outperform non-autoregressive paradigms on structure recovery, while non-autoregressive methods perform better on sequence recovery, as shown in Table~\ref{tab:protein}.
More results are in Appendix.~\ref{app:proteinresults}. 
However, structure recovery is the more appropriate metric since the aim of biomolecule inverse folding is to design a sequence that can be folded into the desired tertiary structure. 
Although biomolecule folding tools can give accurate tertiary structure prediction for a given sequence, they are still time-consuming, which means the evaluation for structure recovery is more time-consuming than sequence recovery. 
We explore the gap between structure recovery (the ratio of TM-score>0.5 \& RMSD<2\AA) and sequence recovery. 
The Pearson correlation coefficient between structure recovery and NSR is 0.6302 and that between structure recovery and perplexity is -0.9029, which indicates that structure recovery and perplexity are highly related, as shown in Fig.~\ref{fig:corr}. The results of structure recovery and sequence recovery are related but not consistent. Therefore, sequence recovery can be utilized as a rough but quick tool for estimating an inverse folding model.

\section{Limitations}

The evaluation relies on existing structure prediction tools (e.g., E2EFold, ESMFold), which may introduce biases or noise in the structural recovery scores. Although stochastic-order generation better captures inter-token dependencies, its computational cost is higher than traditional sequential decoding. While we focus on structure recovery, further exploration of downstream biochemical or functional metrics would be needed to fully evaluate semantic fidelity in biological contexts.

\section{Discussion}

Our work rethinks the paradigm of language models in biological systems by analyzing the intrinsic differences between natural sentences and biological sequences. Through extensive experiments on RNA and protein inverse folding, we demonstrate that stochastic-order decoding significantly improves both sequence and structure recovery, validating our hypothesis that biological languages require generation paradigms beyond left-to-right autoregression. Furthermore, we find that traditional NLP metrics such as BLEU or perplexity are not sufficient for evaluating semantic consistency in biological sequences, and propose a comprehensive evaluation pipeline that integrates structural metrics like TM-score and RMSD. Interestingly, our results highlight that high sequence recovery does not necessarily indicate high structural fidelity, which challenges the assumptions underlying many existing benchmarks. This suggests a need for a paradigm shift in both model design and evaluation from NLP to biological language models.

\bibliography{acl_latex}

\clearpage
\appendix

\newpage
\section{Appendix}
\subsection{Method Details}
\subsubsection{Beam Search}
The process of beam search is described in Algorithm~\ref{alg:bs1} and Algorithm~\ref{alg:bs2}. The purpose of beam search is to extend the search space the sample process and limit the complexity of the algorithm at the same time. We have two types of beam search, \ie, \textbf{decode position based} and \textbf{decode type based}. 
The decode position-based beam search algorithm starts from different positions to begin our stochastic autoregressive decoding process. For we always choose the decode position with the highest confidence, we will choose top $w$, \ie, beam width, the most confident decoded positions, as the start decode position. Finally, we calculate the average confidence of $w$ decoded sequences and select the best one as the final sample result. The decode type-based beam search algorithm starts from the same decode position but uses different acid types. Subsequently, we select the sequence with the highest confidence from all candidate sequences. Typically, the beam width $w$ is set as the number of candidate types, \ie, four for ribonucleic acids in RNA inverse folding task, to acquire the best performance.

\begin{algorithm*}[h!]
	\caption{Decode position-based beam search} 
 	\renewcommand{\algorithmicrequire}{\textbf{Input:}}
	\renewcommand{\algorithmicensure}{\textbf{Output:}}
	\label{alg:bs1} 
	\begin{algorithmic}
		\REQUIRE $\text{sequence length } N, \text{latent vector } \mathbf{h_V}, \text{beam width } w$ 
            \STATE $i \gets 1$, $\mathbf{h_S\gets0}$
            \STATE $\text{decoded position}\gets \emptyset$, $\text{start positions} \gets \emptyset$, candidate sequence $\gets\emptyset$
            \STATE $\text{probs}\gets \text{decoder}(\mathbf{h_V,h_S})$
            \STATE start position $\{a_1,a_2,\cdots,a_w\}\gets$ top $w$ highest probs' position
            \FOR{$j=1$ to $w$}
            \STATE Add $a_j$ into decoded position
            \REPEAT
            \STATE Update $\mathbf{h_S}$ according to decoded position
            \STATE $\text{probs}\gets \text{decoder}(\mathbf{h_V,h_S})$
            \STATE $a_\text{max}\gets \text{arg}\max{(\text{probs})}$
            \STATE Add $a_\text{max}$ into decoded position
            \STATE $i\gets i+1$
            \UNTIL{$ i=N $}
            \STATE finish decoding $s_j$
            \STATE Add $s_j$ into candidate sequence
            \STATE decoded position $\gets \emptyset$, $i\gets 1$
            \ENDFOR
            \STATE $s\gets$select best candidate sequence from $\{s_1,\cdots,s_w\}$
		\ENSURE $\text{decoded sequence } s $ 
	\end{algorithmic} 
\end{algorithm*}

\begin{algorithm*}[h!]
	\caption{Decode type-based beam search} 
 	\renewcommand{\algorithmicrequire}{\textbf{Input:}}
	\renewcommand{\algorithmicensure}{\textbf{Output:}}
	\label{alg:bs2} 
	\begin{algorithmic}
		\REQUIRE $\text{sequence length } N, \text{latent vector } \mathbf{h_V}, \text{beam width } w$ 
            \STATE $i \gets 1$, $\mathbf{h_S\gets0}$
            \STATE $\text{decoded position}\gets \emptyset$, $\text{start positions} \gets \emptyset$, candidate sequence $\gets\emptyset$
            \STATE $\text{probs}\gets \text{decoder}(\mathbf{h_V,h_S})$
            \STATE $a_\text{max}\gets \text{arg}\max{(\text{probs})}$
            \STATE start type$\gets \{x_1,\cdots,x_w\}$
            \FOR{$j\gets 1$ to $w$}
                \STATE Set the type of $a_\text{max}$ as $x_j$
                \STATE Add $a_\text{max}$ into decoded position
                \REPEAT
                    \STATE Update $\mathbf{h_S}$ according to decoded position
                    \STATE $\text{probs}\gets \text{decoder}(\mathbf{h_V,h_S})$
                    \STATE $a_\text{max}\gets \text{arg}\max{(\text{probs})}$
                    \STATE Add $a_\text{max}$ into decoded position
                    \STATE $i\gets i+1$
                \UNTIL{$i=N$}
                \STATE finish decoding $s_j$
                \STATE Add $s_j$ into candidate sequence
                \STATE decoded position $\gets \emptyset$, $i\gets 1$
            \ENDFOR
	\end{algorithmic} 
\end{algorithm*}

\begin{figure*}[h!]
    \centering
    \includegraphics[width=1\linewidth]{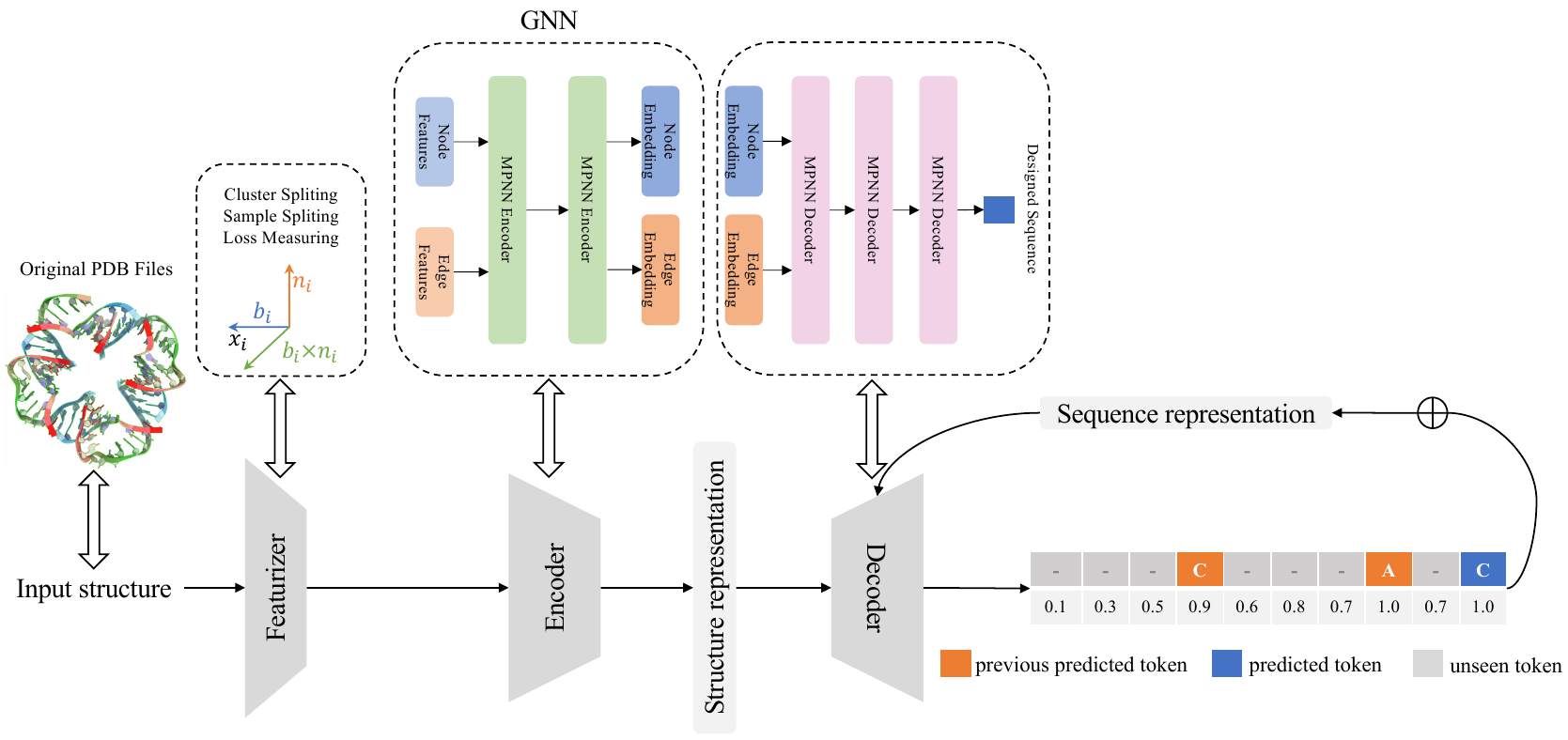}
    \caption{The detail architecture of RiFold.}
    \label{fig:model}
\end{figure*}

\subsubsection{Model architecture}
\label{app:model}
For a fair comparison, we adapt the featurizer of RDesign for RiFold and the featurizer of PiFold for PiFold\_AR. The details of RiFold are described in Fig.~\ref{fig:model}, including the featurizer, encoder, and autoregressive decoder.
Node attributes, denoted as $V \in \mathbb{R}^{N \times f_n}$, comprise $f_n$-dimensional characteristics for $N$ nucleotides that elucidate the local geometric configuration of each nucleotide. These characteristics entail:
\begin{itemize}
    \item Dihedral angles, articulated through sine and cosine functions;
    \item Spatial distances, transcribed into radial basis functions (RBFs) in relation to a reference atom $P_i$;
    \item Directional vectors, deduced in accordance with the local coordinate system $Q_i$.
\end{itemize}

Edge attributes, represented as $E \in \mathbb{R}^{N \times K \times f_m}$, include $f_m$-dimensional characteristics for the $K$ neighbors of each nucleotide, delineating the relative geometric relationships among nucleotides. These characteristics consist of:
\begin{itemize}
    \item Orientation encoding, inferred from the quaternion presentation of the relative rotation between $Q_i$ and $Q_j$;
    \item Spatial distances, transcribed into RBFs among inter-nucleotide atoms and the reference atom $P_i$;
    \item Directional vectors, calculated in relation to the reference atom $P_i$.
\end{itemize}

\subsection{Implementation details}
\label{app:implementation}
\subsubsection{Hyperparameter}
We train all the models for 200 epochs and take the best checkpoint on evaluation. The shown results are on the test set. For RNA inverse folding, we use the optimizer of Adam with a learning rate of 0.001 following \citep{rdesign}. The batch size is 16. For Protein inverse folding, we use the optimizer of Adam with a learning rate of 0.001 and the scheduler of OneCycleLR. The batch size is 8. The number of layers of our RiFold and PiFold-AR is the same as RDesign and PiFold for fair comparison.

\subsubsection{Evaluation details}
We utilize ESMFold\_v1 for protein folding \citep{esmfold} and E2EFold \citep{eM2dRNAs} for RNA folding. For RMSD calculation, we take the $\alpha$ carbon as the backbone for protein and carbon 3 and 4 for RNA. In E2EFold, they relax the predicted structure through a restrained energy minimization procedure as a preventative measure to resolve any remaining structural clashes and violations. Specifically, they minimize the AMBER force field with harmonic restraints, which allows the system to remain close to its input structure. The energy here is taken as our evaluation metric.

\subsubsection{Hardware}
All our experiments are conducted on a computing cluster with 8 GPUs of NVIDIA
GeForce
RTX 4090 24GB and CPUs of AMD
EPYC
7763 64-Core of 3.52GHz. 
All the inferences are conducted on a single GPU of NVIDIA
GeForce
RTX 4090 24GB.


\begin{table}[ht!]
\setlength{\tabcolsep}{2mm}
    \centering
    \begin{tabular}{ccccccc}
    \toprule
        Method &  short &  medium &  long \\
    \midrule
        RDesign & 57.04 & 59.85 & 55.23 \\
       \cellcolor{lm_purple_low} RiFold & \cellcolor{lm_purple_low}59.66 &\cellcolor{lm_purple_low} 63.24 & \cellcolor{lm_purple_low}69.23 \\
    \bottomrule
    \end{tabular}
        \caption{The pair accuracy on RNA solo dataset. ($\% \uparrow$)}
    \label{tab:pair}
\end{table}

\subsection{Additional experimental results}
\subsubsection{RNA inverse folding}
\begin{table*}[ht!]
    \centering
    \begin{tabular}{ccccccccc}
    \toprule
    \multirow{2}{*}{Width} & \multicolumn{4}{c}{Macro-F1 ($\times 100$) $\uparrow$} & \multicolumn{4}{c}{Sequence Recovery (\%) $\uparrow$}\\ \cmidrule(l){2-5} \cmidrule(l){6-9}
    & Short & Medium & Long & All & Short & Medium & Long & All \\
    \midrule
         1&  40.87&  45.85&  42.62&  43.64&  41.90&  45.76&  43.72& 42.86\\
         3&  41.15&  45.97&  42.37&  43.68&  42.07&  45.90&  43.43& 43.02\\
         5&  41.22&  46.36&  42.44&  43.87&  42.45&  46.24&  43.59& 43.33\\
    \bottomrule
    \end{tabular}
    \caption{Beam search with different width of the beam on RNAsolo dataset}
    \label{tab:beamsear}
\end{table*}

\begin{table*}[ht!]
    \centering
    \begin{tabular}{ccccccccc}
    \toprule
    \multirow{2}{*}{Width} & \multicolumn{4}{c}{Macro-F1 ($\times 100$) $\uparrow$} & \multicolumn{4}{c}{Sequence Recovery (\%) $\uparrow$}\\ \cmidrule(l){2-5} \cmidrule(l){6-9}
    & Short & Medium & Long & All & Short & Medium & Long & All \\
    \midrule
         1&  50.68&  56.29&  59.66&  58.94&  57.07&  56.27&  61.80& 53.52\\
         3&  52.56&  55.85&  60.17&  59.28&  58.46&  55.91&  62.37& 55.00\\
         5&  51.04&  55.64&  59.92&  59.02&  57.07&  55.74&  62.10& 56.67\\
    \bottomrule
    \end{tabular}
    \caption{Beam search with different width of the beam on RNA-puzzles dataset}
    \label{tab:beamsear2}
\end{table*}
\paragraph{Beam search}
\label{app:beamsearch}
Experimental results of a beam search with different widths of beam are shown in Table~\ref{tab:beamsear} and Table~\ref{tab:beamsear2}. With a wider beam, the performance of RiFold increases.
\begin{table*}[ht!]
    \centering

    \begin{tabular}{p{2.2cm}<{\centering}p{2.1cm}<{\centering}p{1.7cm}<{\centering}p{2.4cm}<{\centering}p{1.5cm}<{\centering}p{1cm}<{\centering}}
    \toprule
        \multirow{2}{*}{Method} & \multicolumn{3}{c}{Structure recovery} & \multicolumn{2}{c}{Sequence recovery} \\
        \cmidrule(l){2-4} \cmidrule(l){5-6}
          &  TM-score>0.5 (\%) $\uparrow$&  RMSD<2 (\%) $\uparrow$&  TM-score>0.5 \& RMSD<2 (\%)$\uparrow$&  Perplexity $\downarrow$& NSR (\%)$\uparrow$\\
    \midrule
        GraphTrans&  81.70&  13.39&  13.39&  6.63& 35.82\\
        GCA&  81.41&  14.30&  14.30&  6.05& 37.64\\
        StructGNN&  83.20&  14.83&  14.83&  6.40& 35.91\\
        AlphaDesign&  87.22&  54.69&  54.69&  6.30& 41.31\\
        \cellcolor{lm_purple_low}AR\_7\_3&  \cellcolor{lm_purple_low}88.75&  \cellcolor{lm_purple_low}57.44&  \cellcolor{lm_purple_low}57.44&  \cellcolor{lm_purple_low}5.37& \cellcolor{lm_purple_low}48.49\\
        GVP&  89.19&  57.64&  57.64&  5.36& 39.47\\
        \cellcolor{lm_purple_low}AR\_5\_5&  \cellcolor{lm_purple_low}89.66&  \cellcolor{lm_purple_low}58.35& \cellcolor{lm_purple_low} 58.35& \cellcolor{lm_purple_low} 4.91& \cellcolor{lm_purple_low}51.27\\
        \cellcolor{lm_purple_low}AR\_8\_2&  \cellcolor{lm_purple_low}89.47&  \cellcolor{lm_purple_low}59.07&  \cellcolor{lm_purple_low}59.07&  \cellcolor{lm_purple_low}4.88& \cellcolor{lm_purple_low}51.87\\
        PiFold&  90.93&  59.44&  59.44&  4.58& 52.17\\
        \cellcolor{lm_purple_low}AR\_9\_1&  \cellcolor{lm_purple_low}90.93& \cellcolor{lm_purple_low} 60.53&  \cellcolor{lm_purple_low}60.53&  \cellcolor{lm_purple_low}4.87& \cellcolor{lm_purple_low}52.04\\
        KWDesign&  89.10&  60.59&  60.59&  4.42& 60.13\\
        \cellcolor{lm_purple_low}AR\_6\_4&  \cellcolor{lm_purple_low}89.29&  \cellcolor{lm_purple_low}60.98&  \cellcolor{lm_purple_low}60.98&  \cellcolor{lm_purple_low}4.90& \cellcolor{lm_purple_low}51.41\\
        \cellcolor{lm_purple_low}ProteinMPNN&  \cellcolor{lm_purple_low}90.88&  \cellcolor{lm_purple_low}61.04&  \cellcolor{lm_purple_low}61.04& \cellcolor{lm_purple_low} 4.61& \cellcolor{lm_purple_low}45.96\\
    \bottomrule
    \end{tabular}
        \caption{The overall sequence recovery and structure recovery of proteins on the CATH dataset.}
    \label{tab:protein_all}
\end{table*}

\subsubsection{Protein inverse folding}
\label{app:proteinresults}
\begin{figure*}[ht!]
    \centering
    \begin{subfigure}[t]{0.48\textwidth}
        \centering
        \includegraphics[width=\linewidth]{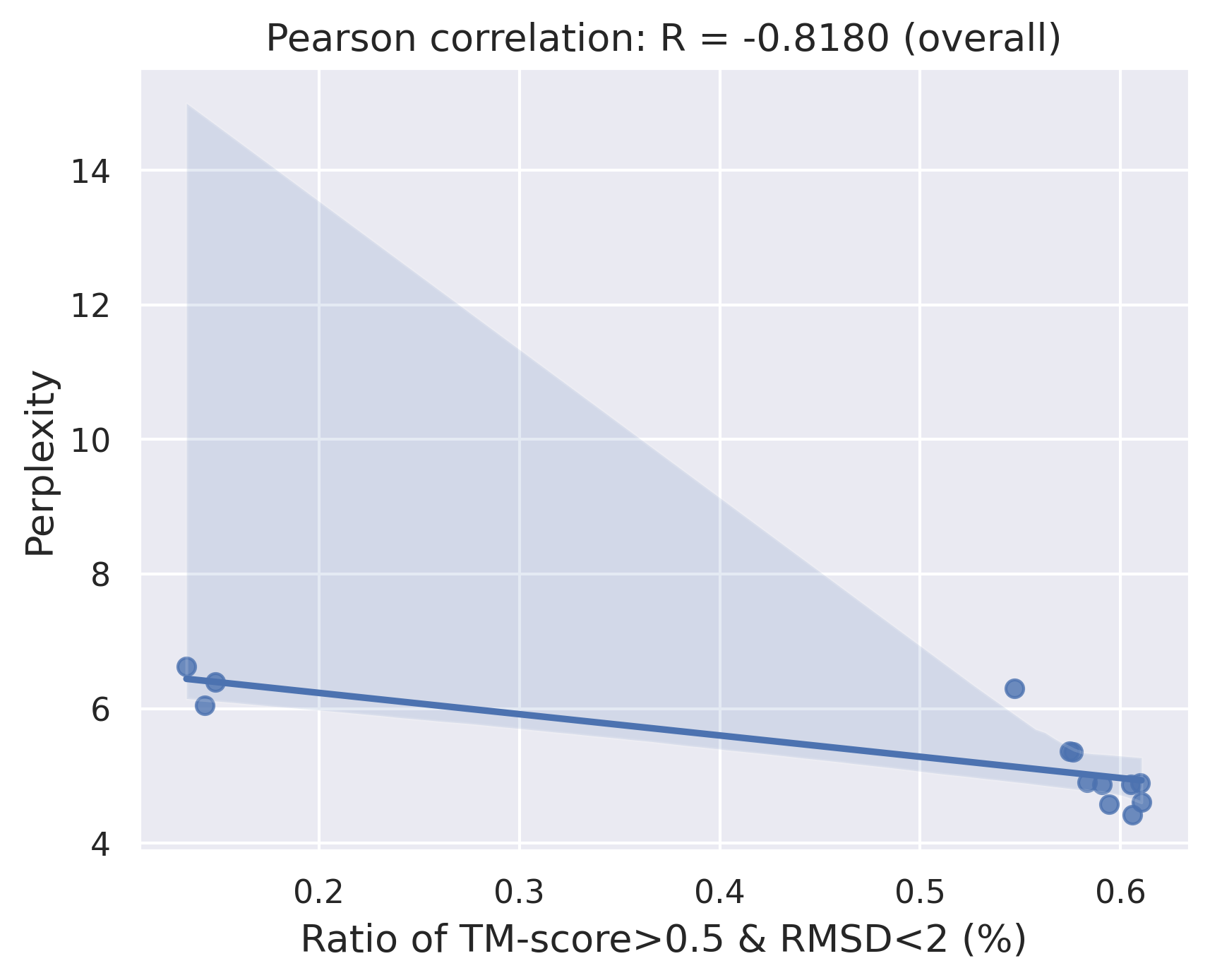}
        \caption{PPL Correlation}
        \label{fig:ppl}
    \end{subfigure}
    \hfill
    \begin{subfigure}[t]{0.48\textwidth}
        \centering
        \includegraphics[width=\linewidth]{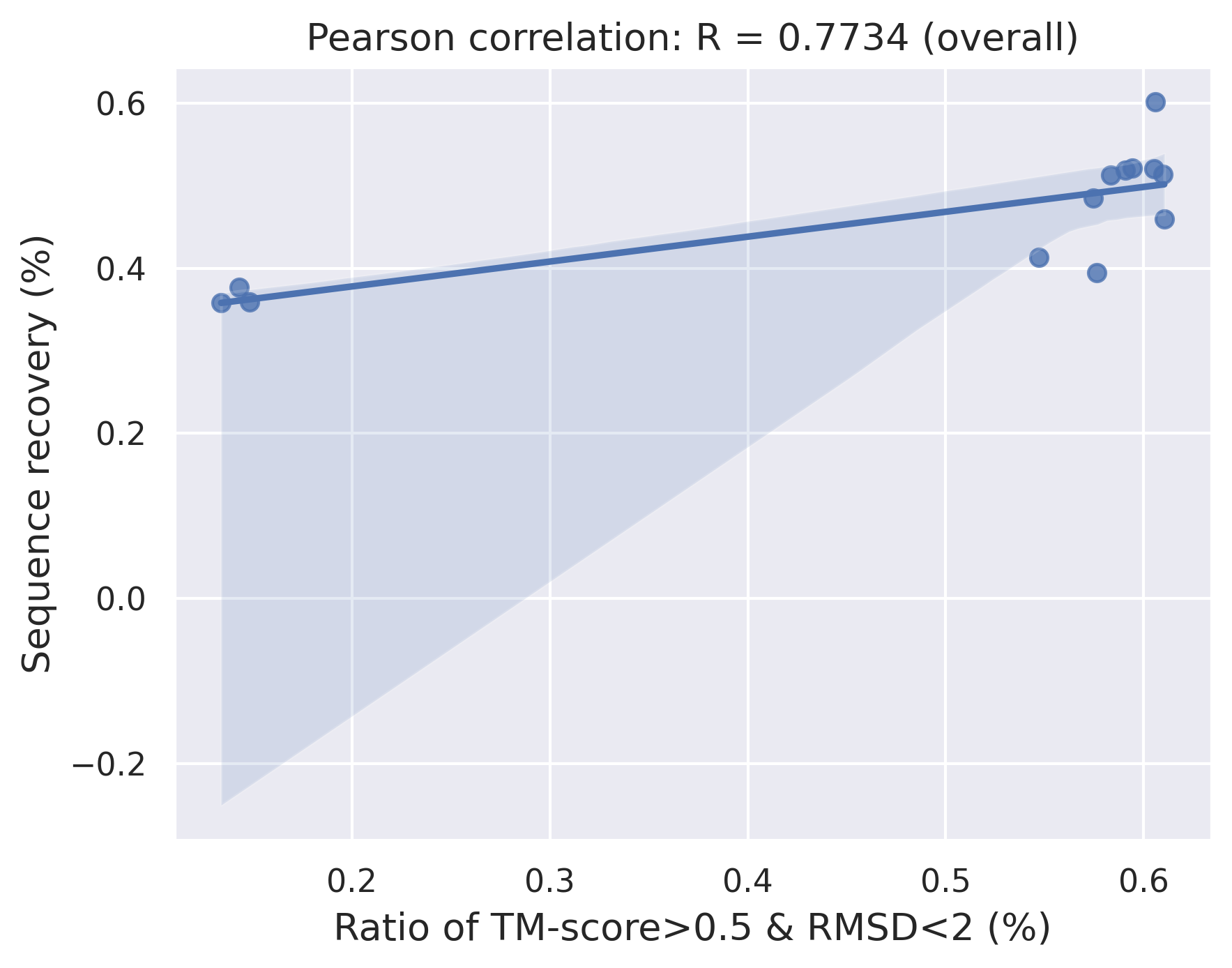}
        \caption{Sequence Recovery}
        \label{fig:seq_rec}
    \end{subfigure}
    \caption{The correlation between structure and sequence recovery.}
    \label{fig:corr_all}
\end{figure*}

The overall sequence recovery and structure recovery of proteins on the CATH dataset are shown in Table~\ref{tab:protein_all}, where AR\_\textit{N}\_\textit{M} indicates the model consists of a \textit{N}-layer PiFold encoder and a \textit{M}-layer autoregressive decoder. The overall sequence recovery and structure recovery of proteins on the TS50 and TS500 datasets are shown in Table~\ref{tab:protein_ts50} and Table~\ref{tab:protein_ts500}. We also calculate the correlation between structure recovery and sequence recovery. The results of the correlation between sequence recovery and structure recovery from all experiments are shown in Fig.~\ref{fig:corr_all}. The Pearson correlation between perplexity and structure recovery is -0.8180, and the Pearson correlation between sequence recovery and structure recovery is 0.7734. Since the gap between different models is too large, we also calculate the Pearson correlation among the top 10 models in Table.~\ref{tab:protein_all}. The correlation results are shown in Fig.~\ref{fig:corr}. Among the top-10 models, the Pearson correlation between perplexity and structure recovery is -0.9029, and the Pearson correlation between sequence recovery and structure recovery is 0.6302. Empirical results show that the two recoveries are related but not consistent.

\begin{table*}[ht!]
    \centering

    \begin{tabular}{cccc}
    \toprule
        Method &  TM-score>0.5 (\%) $\uparrow$&  RMSD<2 (\%) $\uparrow$& \cellcolor{lm_purple_low} TM-score>0.5 \& RMSD<2 (\%)$\uparrow$ \\
    \midrule
        PiFold & 93.88 & 71.43 & \cellcolor{lm_purple_low}71.43 \\
        AR\_5\_5 & 89.80 & 79.59 & \cellcolor{lm_purple_low}\textbf{79.59} \\
        AR\_6\_4 & 91.84 & 75.51 & \cellcolor{lm_purple_low}75.51 \\
        AR\_7\_3 & 93.88 & 73.47 & \cellcolor{lm_purple_low}73.47 \\
        AR\_9\_1 & 87.76 & 73.47 & \cellcolor{lm_purple_low}73.47 \\
    \bottomrule
    \end{tabular}
        \caption{The overall sequence recovery of proteins on the TS50 dataset.}
    \label{tab:protein_ts50}
\end{table*}

\begin{table*}[t!]
    \centering

    \begin{tabular}{cccc}
    \toprule
        Method &  TM-score>0.5 (\%) $\uparrow$&  RMSD<2 (\%) $\uparrow$&  \cellcolor{lm_purple_low}TM-score>0.5 \& RMSD<2 (\%)$\uparrow$ \\
    \midrule
        PiFold & 94.49 & 68.16 & \cellcolor{lm_purple_low}68.16 \\
        AR\_5\_5 & 93.87 & 68.92 & \cellcolor{lm_purple_low}\textbf{68.92} \\
        AR\_6\_4 & 94.08 & 68.37 & \cellcolor{lm_purple_low}68.37 \\
        AR\_7\_3 & 93.88 & 68.98 & \cellcolor{lm_purple_low}68.98 \\
        AR\_9\_1 & 93.87 & 67.89 & \cellcolor{lm_purple_low}67.89 \\
    \bottomrule
    \end{tabular}
        \caption{The overall sequence recovery of proteins on TS500 dataset.}
    \label{tab:protein_ts500}
\end{table*}

\subsubsection{Pair correlation}
\label{app:pair}
We also calculate the accuracy of pairs in RNA. The results are shown in Table~\ref{tab:pair}. For each pair in RNA, \ie, [A,U] and [C,G] in the ground truth, we calculate the accuracy of the acid of the pairs. Our RiFold outperforms RDesign, especially on the long and medium RNAs. The better performance of RiFold on long RNAs comes from the long-context correlation of RNAs. This is caused by the problem that one tertiary structure corresponds to multiple sequences, which means RDesign may predict multiple tokens for one position $p(\pa_i|\pX)$. For RiFold, autoregressive methods alleviate the problem by predicting tokens with the knowledge of known tokens $p(\pa_i|\pa_{\rm known}, \pX)$. More results are in Appendix.~\ref{app:pair}.

\end{document}